1# Automated volumetric and statistical shape assessment of cam-type morphology of the femoral head-neck region from 3D magnetic resonance images

Jessica M. Bugeja[a,b,*], Ying Xia[b], Shekhar S. Chandra[a], Nicholas J. Murphy[c,d], Jillian Eyles[c,e], Libby Spiers[f], Stuart Crozier[a], David J. Hunter[c,e], Jurgen Fripp[b], Craig Engstrom[g]

[a]School of Information Technology and Electrical Engineering, The University of Queensland, Australia

[b]Australian e-Health Research Centre, CSIRO Health and Biosecurity, Australia.

[c]Kolling Institute of Medical Research, Institute of Bone and Joint Research, University of Sydney, Australia

[d]Department of Orthopaedic Surgery, John Hunter Hospital, Newcastle, Australia

[e]Department of Rheumatology, Royal North Shore Hospital, St Leonards, Australia

[f]Centre for Health, Exercise and Sports Medicine, Department of Physiotherapy, University of Melbourne, Melbourne, Australia

[g]School of Human Movement Studies, The University of Queensland, Australia

*Corresponding Author## Abstract

**Background:** Femoroacetabular impingement (FAI) cam morphology is routinely assessed using two-dimensional alpha angles which do not provide specific data on cam size characteristics.

**Purpose:** To implement a novel, automated three-dimensional (3D) pipeline, CamMorph, for segmentation and measurement of cam volume, surface area and height from magnetic resonance (MR) images in patients with FAI.

**Study Type**: Cohort

**Subjects**: A total of 97 patients with FAI (56 males, 41 females) from the Australian FASHIoN trial.

**Sequence**: 3D T2-weighted true fast imaging with steady-state precession.

**Assessment**: The CamMorph pipeline involves two processes: i) proximal femur segmentation using an approach integrating 3D U-net with focused shape modelling (FSM); ii) use of patient-specific anatomical information from 3D FSM to simulate 'healthy' femoral bone models and pathological region constraints to identify cam bone mass. Agreement between manual and automated segmentation of the proximal femur was evaluated with the Dice similarity index (DSI) and surface distance measures.

**Statistical Tests**: Independent t-tests or Mann-Whitney U rank tests were used to compare the femoral head volume, cam volume, surface area and height data between female and male patients with FAI.

**Results**: There was a mean DSI value of 0.964 between manual and automated segmentation of proximal femur volume. Compared to female FAI patients, male patients had a significantly larger mean femoral head volume (66.12cm$^3$ v 46.02cm$^3$, *p*<0.001). Compared to female FAI patients, male patients had a significantly larger mean cam volume (1136.87mm$^3$ v 337.86mm$^3$, *p*<0.001), surface area (657.36mm$^2$ v 306.93mm$^2$, *p*<0.001), maximum-height (3.89mm v 2.23mm, *p*<0.001) and average-height (1.94mm v 1.00mm, *p*<0.001).

**Data Conclusion**: Automated analyses of 3D MR images from patients with FAI using the CamMorph pipeline showed that, in comparison with female patients, male patients had significantly greater cam volume, surface area and height.

**Abbreviations**
3D Three-dimensional
MR Magnetic resonance
FAI Femoroacetabular impingement
true-FISP true fast imaging with steady-state precession
DESS Double echo steady state
ROI Region of interest
PT Physiotherapist-led conservative intervention
DSI Dice similarity index
ASD Average surface distance
HD Hausdorff distance

# Introduction

Cam-type femoroacetabular impingement (FAI) syndrome is characterised by an abnormal asphericity of the femoral head which predisposes to impingement at the anterosuperior acetabular rim and is frequently associated with pain during manoeuvres involving internal hip rotation [1,2]. In clinical investigations, cam morphology is commonly assessed using techniques including computerised tomography (CT) or magnetic resonance (MR) imaging whereby a planar projection, from anterosuperior and/or anterior slices of the femoral head-neck junction, is used to calculate a two-dimensional (2D) alpha angle to appraise femoral head asphericity in patients with suspected FAI [3,4]. An alpha angle of greater than 55° is routinely considered indicative of a cam formation in the femoral head-neck region although higher thresholds have been proposed in relation to increased specificity of symptomatic FAI [5-7]. Currently, the alpha angle is the most commonly used measure of cam morphology in hip arthroscopic surgery [8], although it has moderate to poor inter- and intra-reader reliabilities [9] and these 2D measurements do not provide geometric data on cam size (e.g. volume, surface area, height) for evaluating the specific morphometric characteristics of these osseous formations.

The development of a fully automated 3D approach for the assessment of cam morphology, particularly from clinical MR images, to provide fast, reliable analyses using powerful deep learning methods has the potential to improve pre- and post-intervention quantitative assessment in patients with FAI [10]. The aim of the current work is to develop CamMorph, a novel, fully automated pipeline integrating 3D U-net [11] with focused shape modelling (FSM) based [12] segmentations of the proximal femur for clinical MR images of the hip in patients with symptomatic FAI to calculate femoral head volume, cam volume, surface area and height. This approach combines U-net's ability to localize both high and low-level features in MR images with FSM to generate an anatomical shape *prior* for creating patient-specific 'healthy' models from bone models with cam morphology to provide a robust pipeline for measuring cam volume, surface area and height.

In the present study, the integrated U-net+FSM pipeline was developed to address the challenging task of automated segmentation of cam morphology from clinical 3D MR images in patients with differing FAI severity and pathoanatomical characteristics. The automated segmentation of the clinical MR images, which had varying imaging quality in terms of the coverage of the proximal femur, image artefacts and contrast characteristics, was used to calculate cam volume, surface area, maximum-height and average-height in male and female patients with FAI from the Australian FASHIoN trial [13].

## Materials and Methods

In the current study, three datasets of hip MR images were used including two (i.e., Dataset A [14] and B [15]) for model training and algorithm development and one clinical dataset (Dataset C) [13] for method evaluation.

### Training Datasets

Dataset A contained true fast imaging with steady-state precession (true-FISP) unilateral hip images with manual bone labelling for the proximal femur and acetabulum from a previous study on automated hip joint segmentation in 56 asymptomatic volunteers [14]. This dataset was used for the training of the 3D U-net networks to segment the proximal femur volume specifically for 3D true-FISP images of the hip joint.

Dataset B included 3D bone surfaces of the proximal femur that were previously reconstructed from bilateral water-excited 3D dual echo steady state (DESS) hip MR images for statistical shape modelling [15]. In this work, the training bone surfaces were used to create the FSM with the focus on the femoral head-neck area, targeting shape assessment of cam morphology in this region.

### Testing Dataset

The Australian FASHIoN trial MR dataset (Dataset C) contained unilateral images of the hip from 97 patients with FAI (56M, 41F, aged 16 – 63 years) with a baseline examination of the affected joint. All patients recruited for the Australian FASHIoN trial had hip pain, radiographic signs of cam and/or pincer morphology and had a surgical opinion that the patient would benefit from arthroscopic hip surgery (for further information on patient recruitment, inclusion and exclusion criteria refer to Murphy *et al.'s* paper [13]). The baseline MR images, which were obtained from patients randomly allocated to either a physiotherapist-led conservative intervention (PT) or hip arthroscopy management arm, were acquired with a 3D T2-weighted true-FISP sequence (Repetition time: 10.2ms, echo time: 4.3ms, weighted average image spacing (voxel size varied in this multi-site study): 0.644 x 0.644 x 0.668mm$^3$, field of view (FOV): 16 x 16cm, matrix size: 256 x 256).

**Proximal Femur Volume Segmentation**

To quantitatively evaluate the segmentation performance of the proposed pipeline for an independent dataset of clinical MR images, manual segmentations encompassing the femoral head, neck and proximal shaft were obtained from images in a randomly chosen subset of the PT (N=22) and arthroscopy patients (N=22) in the FASHIoN dataset. The segmentations were performed by JB (under expert guidance from CE, an experienced analyst with 20 years of experience).

**CamMorph (U-net+FSM) Pipeline**

An overview of the CamMorph pipeline developed in the present study is shown in Fig. 1. The three central components involved in the development of the automated segmentation and volumetric analysis framework are: component 1: an initial proximal femur bone 3D U-net segmentation computed on unilateral 3D true-FISP MR images (Fig. 1a), component 2: femoral head segmentation and volume calculation, component 3: cam morphology segmentation for cam volume, surface area and height calculation. The U-net segmentation, Fig. 1b, was used to perform automated bone segmentation of the proximal femur. The second component (Fig. 1c) includes an initial femoral head segmentation with the circle Hough transform and refining of the segmentation using the U-net segmentation. The third component (Fig. 1d-g) starts with a femur bone fitting to the U-net segmentation using femoral bone initialization as per Nishii *et al.* [16], refined with FSM and atlas-based image registration (Fig. 1d). Subsequently, a simulated 'healthy' bone model is created for each femur using FSM methods focused on the femoral head-neck region (see Fig. 1e where red indicates the region of interest (ROI) for cam morphology and blue is outside the ROI for cam morphology) and the position and size of the cam morphology is determined using distance constraints from the simulated 'healthy' bone to the cam morphology in the anterior and anterosuperior quadrant of the femoral head-neck region (Fig. 1f-g). Fig. 1f displays a femoral bone 3D model with signed distance map scalars representative of the distance between the pathological and simulated 'healthy' bone. The distance scalars outside the cam region were thresholded to a value of 0.0. The lower section of Fig. 1f shows the cam region used within this study (where red indicates the ROI and blue is outside this ROI). Fig. 1g shows a flow chart indicating all vertices with a positive distance from the simulated 'healthy' bone (in the cam region) were included as cam morphology. Any vertices without a positive distance were not included as part of the cam bone mass.

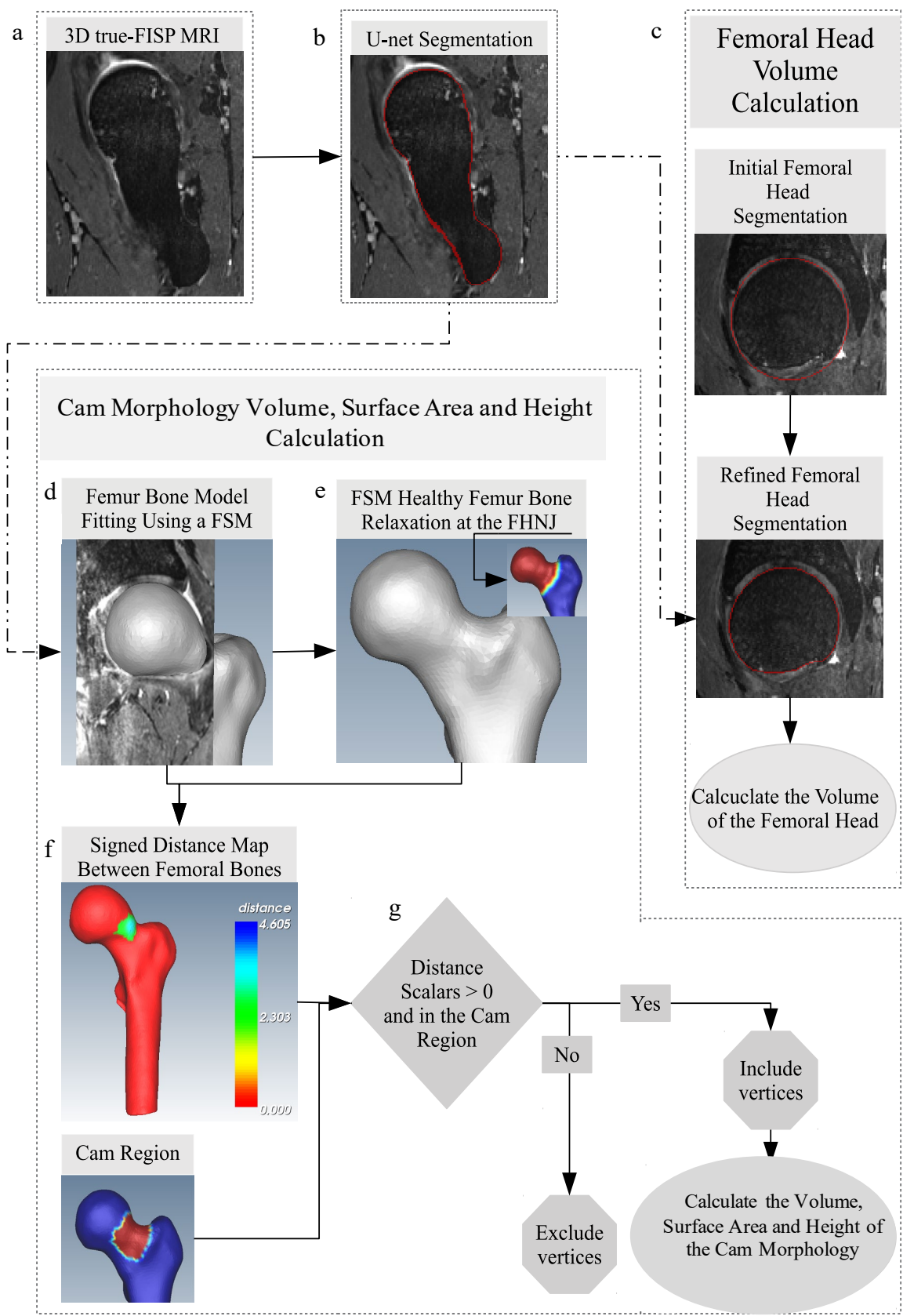

**Figure 1:** CamMorph pipeline for femoral head and cam morphology analysis. (a): Initial T2 weighted 3D true-FISP MR images. (b): U-net femur bone segmentation of FASHIoN MR images. (c): Femoral head segmentation and volume calculation from the FASHIoN data. (d-g): Cam morphology segmentation for calculation of cam volume, surface area and height from the FASHIoN MR dataset. Where FSM = focused shape model and FHNJ = femoral head-neck junction.

## Automated 3D U-net based segmentation of the proximal femoral bone

Automated 3D U-net networks trained on Dataset A [14] were used to perform proximal femur volume segmentation. All MR images from Dataset A were pre-processed with BSpline image resampling (0.5 mm isotropic), joint locator-based ROI extraction, denoising and N4 bias field correction. An isotropic resampling of 0.5mm was chosen based on the high-resolution (0.23x0.23x0.46mm³) of the unilateral true-FISP images within Dataset A.

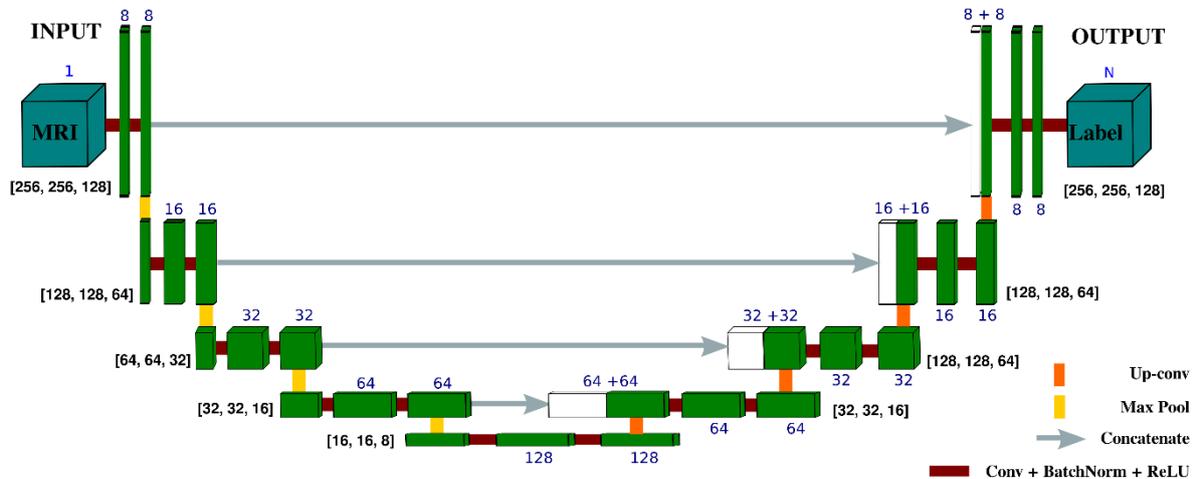

**Figure 2:** U-net architecture. Each green box corresponds to a multi-channel feature map. The number of channels is denoted on top of the box. The x-y-z-size is provided at the lower left edge of the box. White boxes represent copied feature maps. The grey arrows denote concatenate operations, orange boxes correspond to up-convolution operations, yellow boxes denote max pool operations and brown boxes represent a multiple operations step of convolution, batch normalization and rectified linear unit.

## 3D U-net Architecture

The network architecture used in the present work was based on a 3D U-net consisting of a down-sampling path and a symmetric up-sampling modified from its original form [11] for the hip joint (Fig 2). The down-sampling path consists of 5 convolutional blocks, which increases the number of feature channels from 1 to 128. Each down-sampling block has two convolutional layers (kernel size of 3x3x3, stride of 1x1x1) and a Maxpooling layer (pool size of 2x2x2). The fifth block without the Maxpooling layer acts as a bridge to connect both paths. The up-sampling path consists of 4 convolutional blocks, each of which starts with a deconvolutional layer (kernel size of 3x3x3, stride of 2x2x2) and a concatenation with the feature maps corresponding to the down-sampling path. Two convolutional layers are used in each up-sampling block which halve the number of feature channels. All convolutional layers in both down-sampling and up-sampling paths use batch normalization, rectified linear unit activation and 'same padding' (to ensure the output image size is the same as the input image size). In the last layer of the up-sampling path, a 1x1x1 convolution with the Softmax activation function is used to map the multi-channel feature maps to the number of label classes N (N=3; background, femur and acetabulum).

**Training and Segmentation**

All images with manual labels from Dataset A were pre-processed and cropped into an image size of 256x256x128. A 7-fold cross-validation procedure was used where the entire dataset was divided into 7 random subsets. In this procedure, 7 independent training iterations (with 3 repeats) were executed in parallel where each time a different subset of images was used for testing and the other 6 subsets of images for training. For each training iteration, 3D data augmentation (including affine transformation and elastic deformation) was applied to increase the diversity of the training data. Additionally, the Mixup technique was used to reduce the possibility of overfitting and further improve the network performance [17]. During the training process, we used the Dice similarity index (DSI) as the loss function and the adaptive moment estimator (Adam) for parameter optimization (learning rate=0.0001). The entire training procedure resulted in an ensemble of 21 (7x3 repeats) 3D U-net networks trained for segmentation of the femur and acetabulum of the hip joint.

**Segmentation of the FASHIoN Dataset**

All the trained 3D U-net networks were used to automatically segment the bones of the hip joint from MR images in the FASHIoN Dataset, which were independent from the MR data used in the U-net training phase (Dataset A) and FSM (Dataset B). Briefly, to segment a new hip MR image examination, trained 3D U-nets were used to generate a prediction of the femoral and acetabular bones from the pre-processed image. The final bone segmentation was obtained by fusing multiple predicted bone labels via majority voting. In this study, only the femur was used in further processing.

**Femoral Head Segmentation and Volume Calculation**

In the Australian FASHIoN MR dataset, a femoral head segmentation was performed prior to the cam morphology segmentation. This segmentation identified the boundary of the femoral head (which separates the femoral head and neck) within the U-net segmentation (proximal femur volume) and was also used to calculate femoral head volume for statistical comparisons between the male and female patients in the Australian FASHIoN trial. The following text describes the femoral head segmentation pipeline which involved an initial- and refined- segmentation (the latter incorporating the U-net femur bone segmentation).

Fig. 1c illustrates example results from the pipeline used to provide an initial- and refined- segmentation of the femoral head volume from the MR imaging examinations. This initial segmentation involved resampling the MR images to be isotropic before estimations of the femoral head centre and radius were obtained in 3D using the circle Hough transform as per Nishii *et al.* [16]. The refined segmentation used the U-net segmentation to only include voxels within both the U-net- and circle Hough transform-based segmentation. This refined segmentation was required to identify the boundary between the femoral head and neck within the U-net segmentation.

**Cam Segmentation and Measurement Pipeline**

The following section describes the FSM and atlas-based femur bone model initialization, the estimation of a simulated 'healthy' bone model and the identification of cam morphology for calculating cam volume, surface area and height data from the FASHIoN MR Dataset.

**FSM Femur Bone Model Initialization**

Initially, 3D proximal femur bone models were generated using a rough fitting (of the femur bone model to the image) based on the centre of the femoral head [16], followed by FSM-based surface refinement using the U-net segmentations (Fig. 1d) [18]. The ROI on the femur models was defined as shown by the red shading in Fig. 1e. The DSI values of the U-net compared to the manuals were used to identify an atlas set to optimise fitting performance. For cases with DSI<0.95, the initial fitting models were improved using atlas-based registration with the atlas set consisting of cases with DSI≥0.95 fitting accuracy.

**Healthy Femur Bone Generation**

A simulated 'healthy' femur bone model was generated for each patient (Fig. 1e) using a FSM [12], focusing on the femoral head-neck region. This FSM was created using the training surfaces from Dataset B, as per Xia *et al.*'s work [15]. A total of three eigenmodes accounting for 92% of shape variations were visually inspected and used to target the exclusion of any eigenmodes associated with cam morphology from the FSM knowledge.

**Constrained Cam Extent and Location for Cam Morphology Analysis**

To identify the location and quantify the size characteristics of cam morphology, distance-based methods with regional constraints were used. A signed distance map was calculated between the reconstructed proximal femur and the pathological femur surface, using the Visualization Toolkit [19] (VTK) (Fig. 1f). The distance map scalars were overlayed onto the femoral surface for further processing using the simple medical imaging library interface (SMILI) [20].

The anterior and anterosuperior femoral head-neck region (typical location of cam formations) was manually contoured onto the femoral surface (the contouring was performed using SMILI [20]) prior to the initialization of the automatic pipeline. All cases utilized the same contour within the workflow presented. If a vertex on the femoral model was identified to be within this contoured region and the respective femur surface vertex on the distance model had a value of more than 0, the vertex was included as a part of a cam morphology (Fig, 1g). The cam volume and surface area were calculated using VTK [19] and SMILI [20]. The cam height data was obtained at each vertex on the femoral surface using the previously calculated distance map between the simulated 'healthy' femur bone model and the pathological femur model within the cam ROI. The cam maximum- and average-height were calculated from the set of cam height values obtained for each patient.

**CamMorph Evaluation**

**Bone Segmentation and Surface Fitting**

The automated 3D U-net only femur bone segmentations of the MR images from the FASHIoN dataset and FSM fitting to the U-net segmentations were compared against the manual segmentations using DSI, 95% Hausdorff distance (HD) and average surface distance (ASD) values. The equations for these measures are listed below, where *A* denotes the set of automatically segmented voxels and *M* denotes the set of manually segmented voxels:

$$DSI(A, M) = \frac{2|A \cap M|}{|A| + |M|} \tag{1}$$

$$HD(A, M) = \max(h(A, M), h(M, A)) \tag{2}$$

Where *h(A, M)* is the directed HD and is given by:

$$h(A, M) = \max_{a \in A} \min_{m \in M} \|a - m\| \tag{3}$$

The 95th percentile HD is used in this study, instead of the maximum HD (refer to equation 2) due to the sensitivity of this measurement to outliers [21].

$$ASD(A, M) = \max(d(A, M), d(M, A)) \tag{4}$$

Where *d(A, M)* is the directed average HD and is given by:

$$d(A, M) = \frac{1}{N} \sum_{a \in A} \min_{m \in M} \|a - m\|$$

**Statistical Assessment**

Independent t-tests or Mann-Whitney U rank tests were used for analyses between the male and female patients with FAI. Analyses between males and females were completed as compared to females, males have a significantly larger cam morphology size [22,23]. Statistical significance was set at $p<0.05$. Prior to computing the t-tests, the Shapiro-Wilk test was performed to check the normality of the data. Mann-Whitney U rank tests were used if the data did not have a normal distribution. The Levene test was used prior to the t-tests to check for homogeneity of variances. If the input samples did not have equal variances Welch's procedure was applied. The Pearson correlation coefficient was used to assess the association between cam volume and other cam metrics (surface area, maximum- and average-height) in male and female patients. These statistical analyses were performed using a python package, SciPy [24].

# Results

## Manual and Automated Proximal Femur Segmentation

Fig. 3 shows boxplots for the DSI (Fig. 3a), 95% HD (Fig. 3b) and ASD values (Fig. 3c) between the manual and automated bone segmentations of the proximal femur volume obtained from the FASHIoN MR dataset. Overall, the CamMorph (U-net+FSM) pipeline had an enhanced performance compared to U-net alone having better mean scores and decreased spread for the DSI (U-net = 0.958±0.017; CamMorph = 0.964±0.006), 95% HD (U-net = 2.717±2.089mm; CamMorph = 2.123±0.876mm) and ASD values (U-net = 0.577±0.289mm; CamMorph = 0.539±0.189mm).

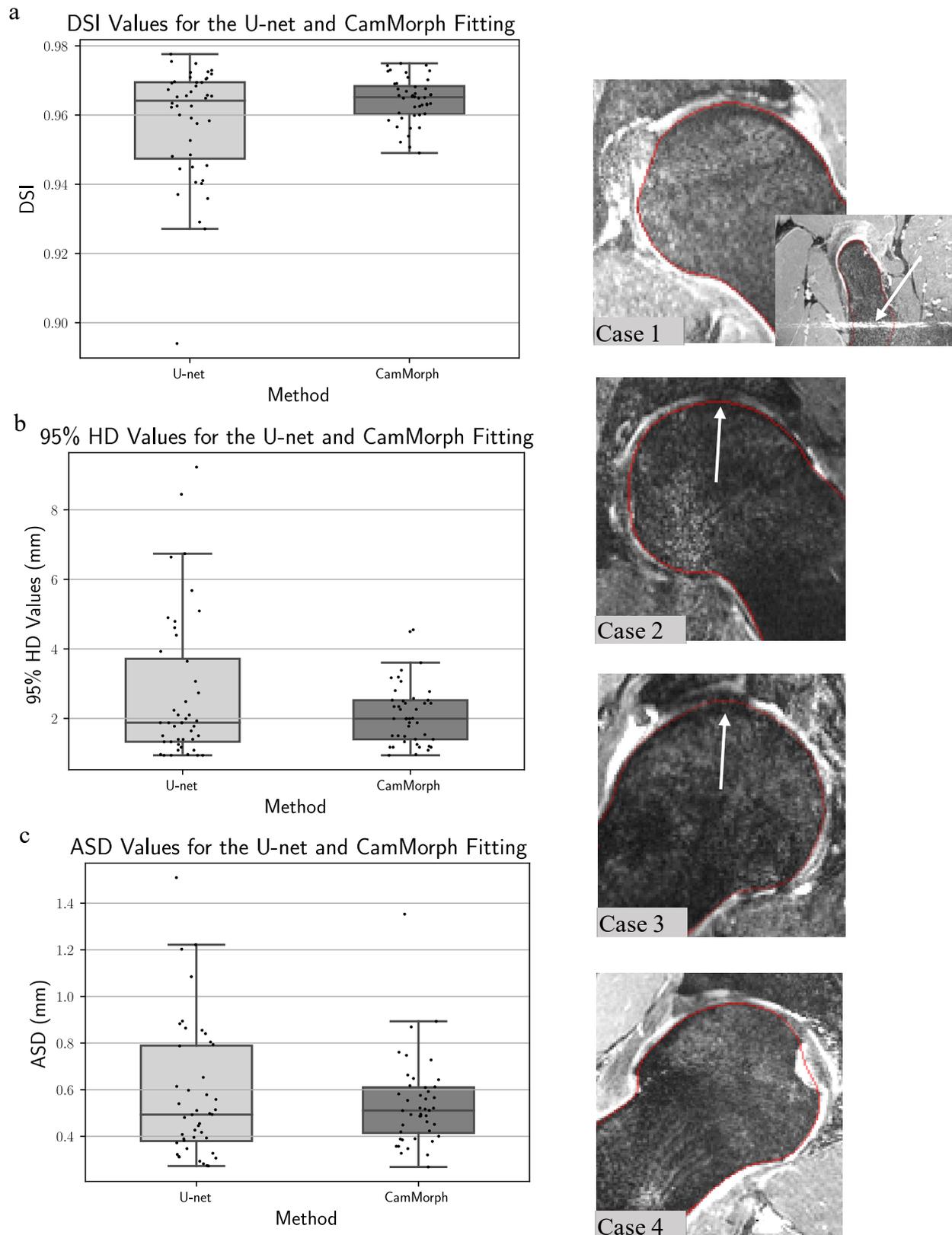

**Figure 3:** Boxplots of the DSI (a), 95% HD (b) and ASD values (c) between the manual and automated segmentations (U-net alone and CamMorph (U-net+FSM)) of the proximal femur (the boxplot centreline marks the median value). The inset coronal MR images show the image quality for outlier cases with the automated segmentation contour overlayed. Case 1 shows nonconforming image contrast characteristics, both Case 1 and Case 2 have motion/image artefacts (Case 1 white arrow in sagittal image indicate artefacts), Case 2 and 3 show comparatively poor bone cartilage interface contrast (white arrows) and Case 4 has, like all the outlier cases shown, incomplete coverage of the femoral neck-greater trochanter region, and a prominent foveal region with surrounding bright synovial fluid which caused difficulty during the FSM fitting. Where DSI = Dice Similarity Index, FSM = Focused Shape Model, HD = Hausdorff Distance, ASD = Average Surface Distance.

Each proximal femur model from the FSM initialization was individually analysed using the HD value. The mean and standard deviation HD between the FSM and the manual segmentation are shown in Fig. 4a and b, respectively. Overall, the model fitting within the femoral head-neck region of the femur, of primary interest in the current study, was in very good agreement with the manual segmentation. Notably, the weighted average image spacing within this study was 0.644x0.644x0.668, whereby on average the highest surface distance accuracy attainable within this study is 0.644mm (the smallest image spacing dimension). Observations from Fig. 4a and b indicate that the HD values within the femoral head-neck region were below 0.644mm in both the mean and standard deviation visualizations, respectively. These results reflect the high surface-based accuracy of the CamMorph pipeline which was robust to the pathologies and image-based issues within the FASHIoN dataset analysed. In the current study, the greater trochanter and the included shaft region of the femur presented with larger HD values, however these are not clinically important areas for cam morphology assessment.

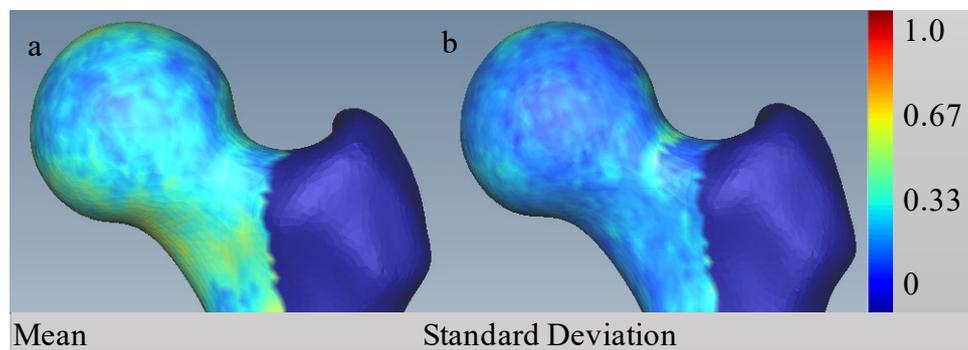

**Figure 4:** Mean (a) and standard deviation (b) Hausdorff distance between the U-net+FSM and the manual segmentations of the proximal femur. The Hausdorff distance scalar bar (rainbow colour map on right side of the figure) has units in mm. The greater trochanter region and femur shaft region (not clinically important areas in this study on cam formations) were given a value of zero for visualization purposes.

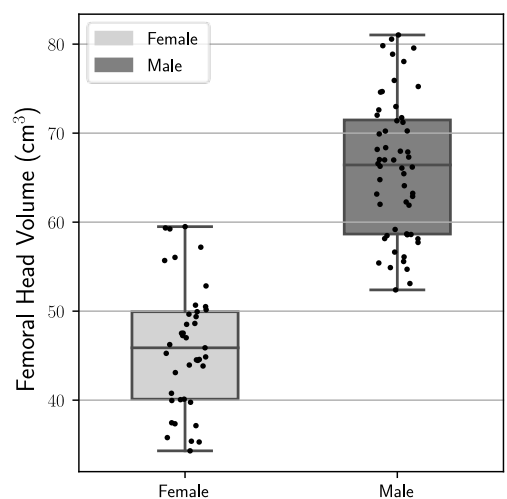

**Figure 5:** Boxplots of femoral head volumes in female and male patients from the FASHIoN MR dataset. The boxplot centreline marks the median value.

**Femoral Head Volume**

Fig. 5 displays the femoral head volumes for the male and female patients from the FASHIoN MR dataset. Compared to female patients, male patients had a significantly larger mean femoral head volume (66.12cm$^3$ v 46.02cm$^3$; t(95)=13.21, $p$<0.001).

**Cam Volume, Surface Area and Height**

Fig. 6 displays boxplots and Table 1 provides the mean and standard deviation data for the cam volume, surface area, maximum-height and average-height for the male and female patients from the FASHIoN MR dataset. Compared to females, males had a significantly larger mean cam volume (1136.87 mm$^3$ v 337.86 mm$^3$), surface area (657.36 mm$^2$ v 306.93mm$^3$), maximum-height (3.89 mm v 2.23 mm) and average-height (1.94 mm v 1.00mm).

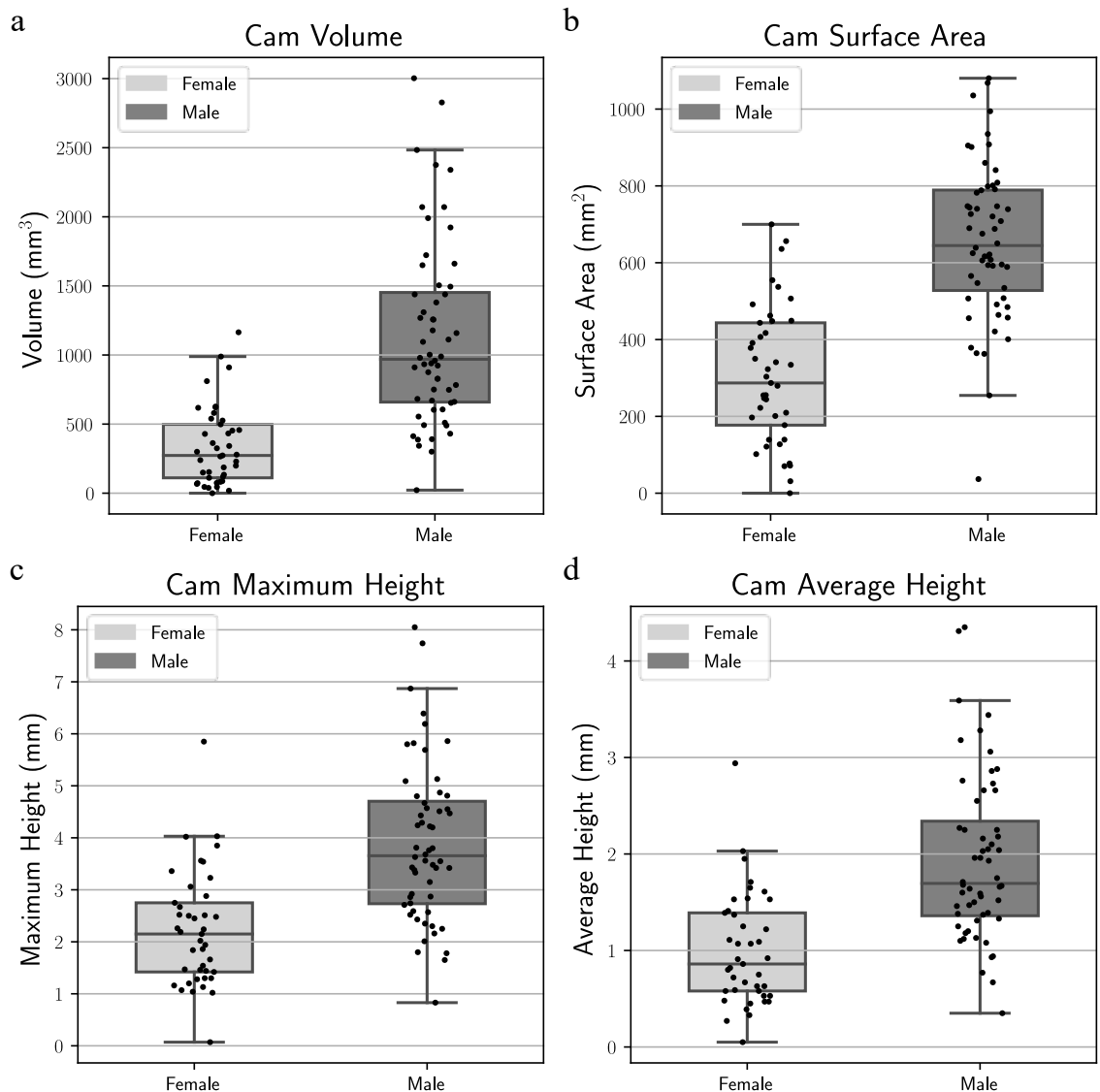

**Figure 6:** Boxplots of cam volume (a), surface area (b), maximum-height (c) and average-height (d) for the female and male patients from the FASHIoN MR dataset. The boxplot centreline marks the median value.

**Table 1**. Mean and standard deviation data for cam volume, surface area, maximum- and average-height in the male and female patients from the FASHIoN MR dataset.

|  | **Male (n=56)** | **Female (n=41)** | **Statistical Significance** |
| --- | --- | --- | --- |
| Volume (mm$^3$) | 1136.87 ± 659.83 | 337.86 ± 279.95 | $U = 240.0, p < 0.001$ |
| Surface Area (mm$^2$) | 657.36 ± 203.04 | 306.93 ± 175.61 | $t(95) = 8.79, p < 0.001$ |
| Maximum Height (mm) | 3.89 ± 1.51 | 2.23 ± 1.09 | $U = 407.0, p < 0.001$ |
| Average Height (mm) | 1.94 ± 0.86 | 1.00 ± 0.57 | $U = 380.0, p < 0.001$ |

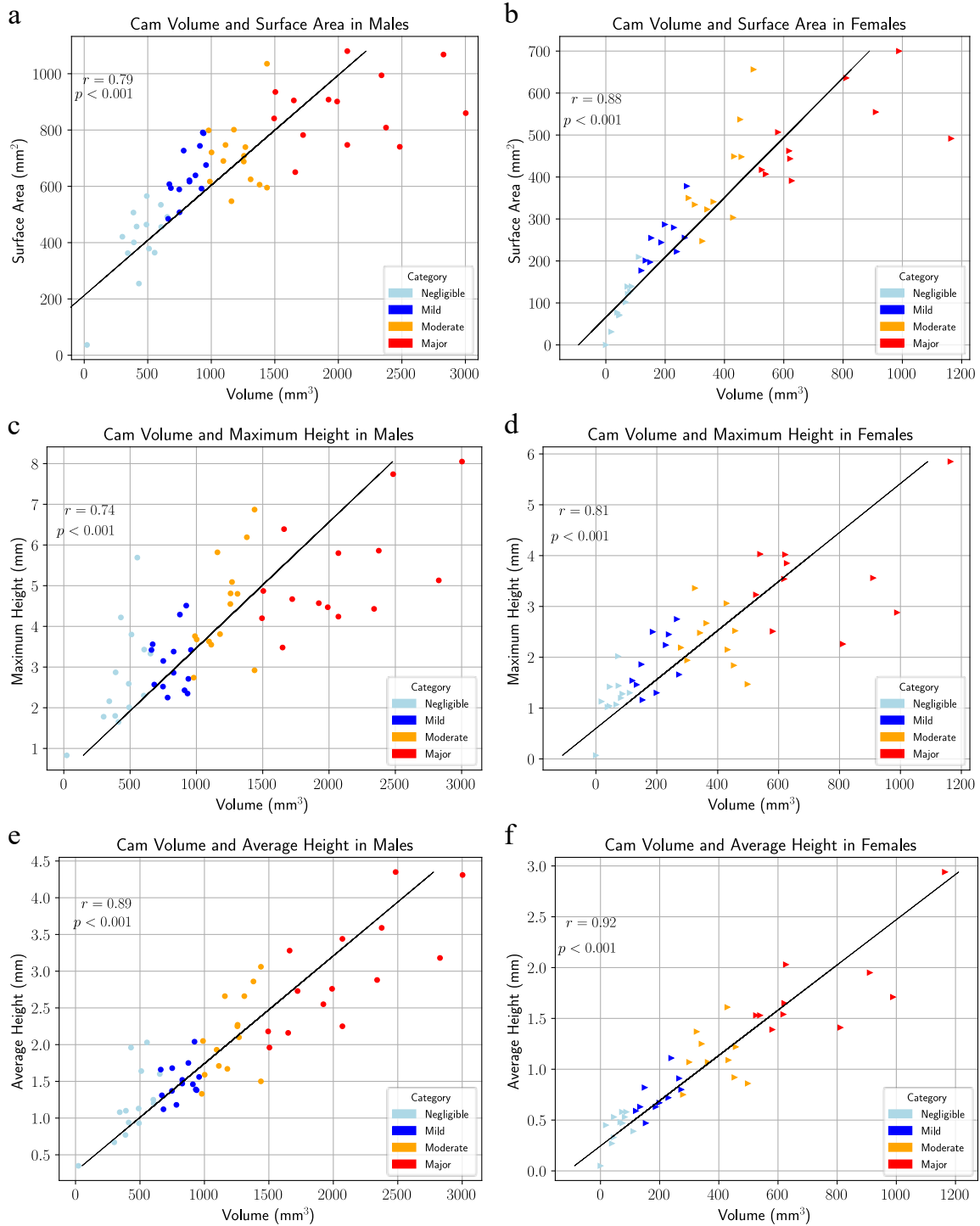

**Figure 7:** Scatterplots (male column 1, female column 2) between cam volume and surface area (a – b), cam volume and maximum-height (c – d) and cam volume and average-height (e – f).

In the male and female patients from the FASHIoN study, cam volume was classified based on quartile data (Table 2), as negligible (≤Q1), mild (>Q1 and ≤Q2), moderate (>Q2 and ≤Q3), or major (>Q3).

**Table 2**. Cam volume quartiles used for classifying cam volume for the male and female patients from the FASHIoN MR dataset.

| Quartile | Male | Female |
|---|---|---|
| Q1 | 657.38 mm$^3$ | 111.06 mm$^3$ |
| Q2 | 969.22 mm$^3$ | 272.97 mm$^3$ |
| Q3 | 1466.51 mm$^3$ | 497.93 mm$^3$ |

For the male and female patients from the FASHIoN MR dataset, there were strong postitive correlations beteween cam volume and surface area (Fig. 7a, b), cam volume and maximum-height (Fig. 7c, d) and cam volume and average-height (Fig. 7e, f).

Fig. 8 shows 3D models of the proximal femur which were generated in SMILI to show the cam profiles in individual female and male patients classified as having a negligible (Fig. 8a, e), minor (Fig. 8b, f), moderate (Fig. 8c, g) or major (Fig. 8d, h) cam volume. Inspection of the cam profiles allowed for comparative visual analysis within and between the classification categories; for example, Fig 8c shows the cam profile in a female patient classified with a moderate cam volume whereby the cam surface area profile is clearly smaller than that in a female patient classified with a minor cam volume (Fig 8b).

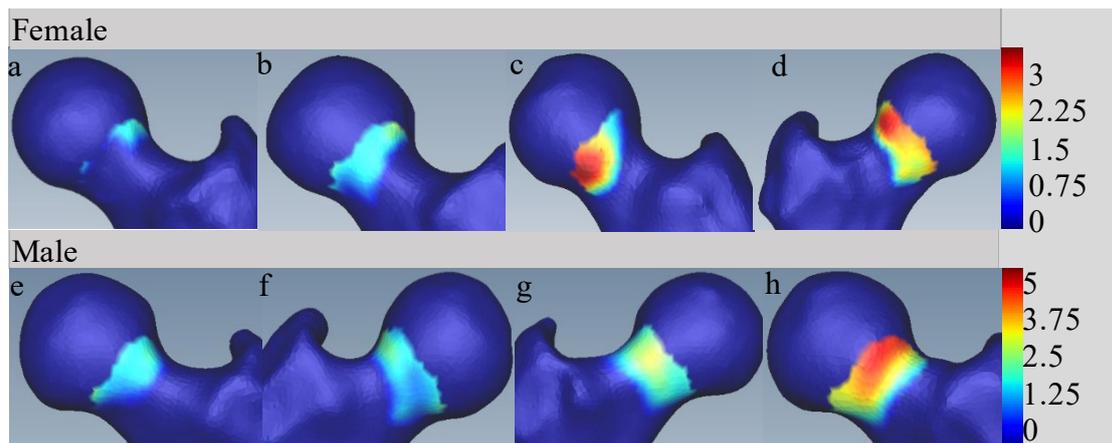

**Figure 8:** Visualization cam profiles in 3D proximal femoral bone models in individual female (a – d) and male (e – h) patients from the FASHIoN MR dataset. The cases displayed include negligible (a and e) mild (b and f), moderate (c and g) and major (d and h) cam volumes. A separate scalar bar (rainbow colour map) for males and females provides cam height in mm.

## Discussion

In the present paper, a novel automated U-net+FSM segmentation and analysis pipeline found cam volume, surface area, maximum-height and average-height were significantly greater in male, compared to female, patients with FAI investigated with 3D MR imaging in the Australian FASHIoN trial. There were significant positive correlations between cam volume and surface area, cam volume and maximum-height and cam volume and average-height in the male and female patients. Generation of 3D bone models of the proximal femur based on segmentations from the automated CamMorph pipeline in FAI patients allowed comparative visualization of cam morphology distributions within and between male and female patients classified with a negligible, minor, moderate or major cam volume.

**Proximal Femur Segmentation**

The high degree of agreement between the manual and CamMorph segmentations of proximal femur volume (mean DSI of 0.964) in the present work compares very favourably with previous studies on automated segmentation of the femur in MR images [25-29] (see Appendix A.1. for specific data [12,26,30-33]). Overall, the combined U-net+FSM approach used in the CamMorph pipeline resulted in a decreased spread of the DSI values (along with an overall increase in the mean value) compared with the U-net alone segmentations of the proximal femur volume. Enhanced performance of the CamMorph pipeline compared with the U-net alone was also found for the 95% HD and ASD data providing very strong support for combining the U-net+FSM approaches for the segmentation of the proximal femur in the clinical MR images obtained from patients with FAI in the Australian FASHIoN study.

Notably, the high surface-based accuracy for segmentation of the proximal femur from clinical MR images by the automated CamMorph pipeline was achieved despite the presence of variable motion artefacts, contrast differences and FOV issues (insufficiencies) within the multi-site FASHIoN MR dataset. For the femoral head-neck region specifically, the majority of the mean 95% HD values between the manual and automated CamMorph segmentations were between 0 to 0.644mm highlighting the robustness of the combined U-net+FSM approach for targeted segmentation of bone in this specific region in the MR images of male and female patients with FAI.

**Femoral Head Volume**

In the present study, the mean femoral head volume in male patients (66.12cm$^3$) was significantly greater than in the female patients (46.02cm$^3$). The femoral head volumes derived from our automated approach are consistent with values reported in earlier CT and MR imaging studies [9,22,23,34-36] (Appendix A.2.). In previous MR arthrography [35], CT [9,22,34] and cadaveric studies [36], femoral head volume has been calculated using manual segmentation [35], region growing methods (based on CT density) [9,22,34] and direct anatomical measurements using callipers [36], whereas there was no manual (or semi-automated) processing required in the current MR-based CamMorph analyses of FAI patients. Furthermore, our automated segmentation of femoral head volume was robust to varying MR image contrast characteristics and artefacts, differing imaging FOV and the presence of osteophytes and increased synovial fluid surrounding the bone-cartilage interface of the femoral head.

**Cam Volume, Surface Area and Height**

The automated CamMorph pipeline presented in the current paper successfully integrated patient-specific healthy models derived from individualised pathological models of the proximal femur for the calculation of cam volume, surface area and height data. Previously, Kang *et al.* [10] proposed calculation of cam morphology with reference to healthy bone models although they used an average model derived from 4 separate femur models from asymptomatic subjects obtained through manual segmentation whereas we used patient-specific healthy models derived from the automated analyses of patients' 3D MR images. The use of the automated 3D CamMorph pipeline to measure cam morphology, as opposed to manual [10] and semi-automated [9,22,23,34,37,38] approaches, offers a fast (within minutes), reliable method to obtain patient-specific data well suited for clinical research as well as larger-scale clinical trials like the Australian FASHIoN study.

In the present analysis of the FASHIoN MR dataset, the mean cam volume in male patients ($1.14cm^3$) was significantly greater than in the female patients ($0.34cm^3$). Recently, Guidetti et al. [39] performed arthroscopic removal of cam bone mass from the hemipelvis of 7 cadavers (sex not specified) and reported that the surgically excised cam volumes, as determined from semi-automated MR measurements, ranged between 0.25 to $1.75cm^3$ which is in good general agreement with our range of cam volumes determined in the male (0 to $3.0cm^3$) and female patients (0 to $1.16cm^3$) using the CamMorph pipeline (Appendix A.3).

Previous CT studies [9,22,34] have reported substantially larger mean cam volumes (e.g., $4.3cm^3$ in females, $6.7cm^3$ in males, 4.6 to $7.96cm^3$ in mixed sex cohorts) from segmentations using a semi-automated region-growing technique based on density (HU) without any reference to the 3D shape profile of the femoral head-neck region. Whilst the substantial differences between the MR derived (current study, Guidetti et al. [48]) and CT derived cam volumes [9,22,34] maybe due to extremely large differences in the severity of the cam lesions in these study cohorts it may relate to an over-estimation of cam volume in the CT studies due to the region-growing approach systematically over-segmenting areas within the lower bone density supero-lateral quadrant of the femoral head [40]. An over-segmentation of cam volume for the above CT studies could account for scenarios such as the relative volume of a cam bone mass ($11.1cm^3$) being 28% of that compared to femoral head volume ($39.1cm^3$) as reported by Schauwecker et al. [34]. In the most extreme cases in the current MR study, the relative cam to femoral head volume was below 5% for the male and female patients with FAI in the Australian FASHIoN study (Appendix A.4).

The greater cam volume in males compared to females observed in the present study is consistent with the findings from previous CT studies [22,23]. Currently, understanding of this sex-based difference in cam volume is incomplete but it may be associated with factors such as males having greater absolute body dimensions, higher magnitude loading of the hip joint or sexual dimorphism in the anatomical and alignment characteristics of the femur and acetabulum. Studies have also reported that cam morphology is more common in males with FAI [41] whereas pincer morphology of the acetabulum is more common in female patients [41] although the presence of both cam and pincer morphology, the so called mixed-type FAI, is common in both male and female patients [41]. In the Australian FASHIoN study, patients were included on the basis of hip pain and radiological signs of FAI related to cam morphology (e.g., alpha angles >55°), pincer morphology (e.g., lateral centre edge angle >40°, positive cross-over sign) and/or mixed-type FAI rather than being solely restricted to those with cam morphology only; this is specifically reflected in the current CamMorph analyses which show both male and female FAI patients with negligible cam volumes.

Compared to female FAI patients, male patients had a significantly larger mean cam surface area ($6.57cm^2$ v $3.07cm^2$), maximum-height (3.89mm v 2.23mm) and average-height

417(1.94mm v 1.00mm) in the FASHIoN MR dataset. These values agree well with the mean cam surface area (6.158cm$^2$), maximum-height (3.7mm) and average-height (1.6 mm) found from the MR measures obtained in the cadaver study of Guidetti *et al.* [39]. Kang *et al.* reported, based on MR measurements, a considerably smaller mean cam surface area (0.52cm$^2$), which may be related to a suboptimal fitting of the average 'healthy' model to the pathological femur bone models [10]. Moreover, their mean cam average-height (3.9mm) was notably different to the mean values obtained from the MR measures in our current work and Guidetti *et al.*'s study. The use of average 'healthy' models by Kang *et al.* [10], rather than an individual patient model as per the current CamMorph approach, may again be related to these differing outcome data.

Strong positive correlations were observed between cam volume and surface area, cam volume and maximum-height and cam volume and average-height in the male and female patients from the FASHIoN study. However, it should be noted that there were clear geometric differences in cam profiles, as visualised in the 3D bone models of the proximal femur in individual patients, within and across patients classified as having negligible, minor, moderate or major cam volumes. Hence, cam morphology does not appear to conform to an underlying geometrically regular shape in the current cohort of patients although further systematic shape based analyses are required to explore this aspect across factors such as sex and FAI severity.

**Future Work**

In the present study, the automated CamMorph pipeline was developed (Datasets A and B) and assessed (FASHIoN dataset) using 3D true-FISP and DESS MR images from three separate databases. In future work, the ability of CamMorph to successfully segment and analyze cam morphology from other 3D MR imaging sequences such as VIBE and SPACE, which have been shown to provide bone volumes comparable to CT in a cadaveric study of the knee joint [42], would be beneficial for further evaluation of this integrated U-net+FSM approach for the analysis of bone formations in symptomatic conditions such as cam-type FAI. Future work involving systematic shape based analyses of cam morphology could be pursued to explore potential associations with clinical and management (e.g., surgery) outcomes collected in the FASHIoN trial or investigate how routine 2D alpha angle data, or alpha angles from multiple sites around the femoral head-neck junction determined from automated analysis of 3D MR images (Xia *et al.* [43]), correlate to cam volume, surface area and height data.

In an upcoming study, we will use the automated CamMorph pipeline for prospective analysis of cam volume, surface area and height data obtained from MR images collected from FAI patients following either conservative PT or arthroscopic surgery (ostectomy). Data derived from the CamMorph pipeline will be used to quantify any management-related change in cam morphology which, in combination with visualisations from the 3D bone models of the proximal femur in individual patients, has potential application for detailed post-treatment (and indeed pre-treatment) analysis of cam morphology in patients with FAI.

In conclusion, the current paper presented a novel automated U-net+FSM segmentation and analysis pipeline, CamMorph, which showed that cam volume, surface area, maximum-height and average-height were significantly greater in male, compared to female, patients with FAI investigated with 3D MR imaging in the Australian FASHIoN trial.

# Conflict of Interest Disclosure

The authors have no COI to report.

# References

1. Ganz R, Parvizi J, Beck M, Leunig M, Nötzli H, Siebenrock KA. Femoroacetabular impingement: a cause for osteoarthritis of the hip. Clinical Orthopaedics and Related Research® 2003;417:112-120.
2. Nötzli H, Wyss T, Stoecklin C, Schmid M, Treiber K, Hodler J. The contour of the femoral head-neck junction as a predictor for the risk of anterior impingement. The Journal of bone and joint surgery British volume 2002;84(4):556-560.
3. Harris MD, Kapron AL, Peters CL, Anderson AE. Correlations between the alpha angle and femoral head asphericity: implications and recommendations for the diagnosis of cam femoroacetabular impingement. European journal of radiology 2014;83(5):788-796.
4. de Sa D, Urquhart N, Philippon M, Ye J-E, Simunovic N, Ayeni OR. Alpha angle correction in femoroacetabular impingement. Knee Surgery, Sports Traumatology, Arthroscopy 2014;22(4):812-821.
5. Sutter R, Dietrich TJ, Zingg PO, Pfirrmann CW. How useful is the alpha angle for discriminating between symptomatic patients with cam-type femoroacetabular impingement and asymptomatic volunteers? Radiology 2012;264(2):514-521.
6. Aliprandi A, Di Pietto F, Minafra P, Zappia M, Pozza S, Sconfienza L. Femoro-acetabular impingement: what the general radiologist should know. Official Journal of the Italian Society of Medical Radiology 2014;119(2):103-112.
7. Masjedi M, Marquardt CS, Drummond IM, Harris SJ, Cobb JP. Cam type femoro-acetabular impingement: quantifying the diagnosis using three dimensional head-neck ratios. Skeletal radiology 2013;42(3):329-333.
8. Audenaert E, Baelde N, Huysse W, Vigneron L, Pattyn C. Development of a three-dimensional detection method of cam deformities in femoroacetabular impingement. Journal of the International Skeletal Society A Journal of Radiology, Pathology and Orthopedics 2011;40(7):921-927.
9. Dessouky R, Chhabra A, Zhang L, et al. Cam-type femoroacetabular impingement—correlations between alpha angle versus volumetric measurements and surgical findings. European Radiology 2019;29(7):3431-3440.
10. Kang X, Zhang H, Garbuz D, Wilson DR, Hodgson AJ. Preliminary evaluation of an MRI-based technique for displaying and quantifying bony deformities in cam-type femoroacetabular impingement. International journal of computer assisted radiology and surgery 2013;8(6):967-975.
11. Çiçek Ö, Abdulkadir A, Lienkamp SS, Brox T, Ronneberger O. 3D U-Net: Learning Dense Volumetric Segmentation from Sparse Annotation. Medical Image Computing and Computer-Assisted Intervention – MICCAI 2016. Cham: Springer International Publishing; 2016. p. 424-432.
12. Chandra SS, Xia Y, Engstrom C, Crozier S, Schwarz R, Fripp J. Focused shape models for hip joint segmentation in 3D magnetic resonance images. Medical Image Analysis 2014;18(3):12.
13. Murphy NJ, Eyles J, Bennell KL, et al. Protocol for a multi-centre randomised controlled trial comparing arthroscopic hip surgery to physiotherapy-led care for femoroacetabular impingement (FAI): the Australian FASHIoN trial. BMC Musculoskeletal Disorders 2017;18(1):406.
14. Xia Y, Chandra SS, Engstrom C, Strudwick MW, Crozier S, Fripp J. Automatic hip cartilage segmentation from 3d mr images using arc-weighted graph searching. Physics in Medicine and Biology 2014;59(23):7245-7266.

15. Xia Y, Fripp J, Chandra SS, Schwarz R, Engstrom C, Crozier S. Automated bone segmentation from large field of view 3d mr images of the hip joint. Physics in Medicine and Biology 2013;58(20):7375-7390.
16. Nishii T, Sugano N, Sato Y, Tanaka H, Miki H, Yoshikawa H. Three-dimensional distribution of acetabular cartilage thickness in patients with hip dysplasia: a fully automated computational analysis of MR imaging. Osteoarthritis and Cartilage 2004;12(8):650-657.
17. Zhang H, Cisse M, Dauphin YN, Lopez-Paz D. mixup: Beyond Empirical Risk Minimization. 2017.
18. Cootes TF, Taylor CJ, Cooper DH, Graham J. Active Shape Models-Their Training and Application. Computer Vision and Image Understanding 1995;61(1):38-59.
19. Schroeder W, Martin K, Lorensen B. The visualization toolkit, 4th edn. Kitware. New York 2006.
20. Chandra SS, Dowling JA, Engstrom C, et al. A lightweight rapid application development framework for biomedical image analysis. Computer Methods and Programs in Biomedicine 2018;164:193-205.
21. Litjens G, Toth R, van de Ven W, et al. Evaluation of prostate segmentation algorithms for MRI: the PROMISE12 challenge. Medical image analysis 2014;18(2):359-373.
22. Zhang L, Wells J, Dessouky R, et al. 3D CT Segmentation of CAM type Femoroacetabular Impingemnet - Reliability and Relationship of CAM lesion with Anthropomorphic Features. Department of Radiology Research Day 2017. Dallas; 2017.
23. Yanke AB, Khair MM, Stanley R, et al. Sex Differences in Patients With CAM Deformities With Femoroacetabular Impingement: 3-Dimensional Computed Tomographic Quantification. Arthroscopy-the Journal of Arthroscopic and Related Surgery 2015;31(12):2301-2306.
24. Virtanen P, Gommers R, Oliphant TE, et al. SciPy 1.0: fundamental algorithms for scientific computing in Python. Nature Methods 2020;17(3):261-272.
25. Zeng G, Wang Q, Lerch T, et al. Latent3DU-net: Multi-level Latent Shape Space Constrained 3D U-net for Automatic Segmentation of the Proximal Femur from Radial MRI of the Hip. In: Shi Y, Suk H-I, Liu M, editors. Machine Learning in Medical Imaging. Cham: Springer International Publishing; 2018. p. 188-196.
26. Zeng G, Zheng G. Deep Learning-Based Automatic Segmentation of the Proximal Femur from MR Images. In: Zheng G, Tian W, Zhuang X, editors. Intelligent Orthopaedics: Artificial Intelligence and Smart Image-guided Technology for Orthopaedics. Singapore: Springer Singapore; 2018. p. 73-79.
27. Peng B, Guo Z, Zhu X, Ikeda S, Tsunoda S. Semantic Segmentation of Femur Bone from MRI Images of Patients with Hematologic Malignancies. 2020 IEEE REGION 10 CONFERENCE (TENCON); 2020. p. 1090-1094.
28. Memiş A, Varlı S, Bilgili F. Semantic segmentation of the multiform proximal femur and femoral head bones with the deep convolutional neural networks in low quality MRI sections acquired in different MRI protocols. Computerized Medical Imaging and Graphics 2020;81:101715.
29. Deniz CM, Xiang S, Hallyburton RS, et al. Segmentation of the Proximal Femur from MR Images using Deep Convolutional Neural Networks. Scientific Reports 2018;8(1):16485.
30. Zeng G, Yang X, Li J, Yu L, Heng P-A, Zheng G. 3D U-net with Multi-level Deep Supervision: Fully Automatic Segmentation of Proximal Femur in 3D MR Images. Machine Learning in Medical Imaging. Cham: Springer International Publishing; 2017. p. 274-282.
31. Zeng G, Zheng G. 3D Tiled Convolution for Effective Segmentation of Volumetric Medical Images. Medical Image Computing and Computer Assisted Intervention – MICCAI 2019. Cham: Springer International Publishing; 2019. p. 146-154.
32. Zeng G, Zheng G. Holistic decomposition convolution for effective semantic segmentation of medical volume images. Medical Image Analysis 2019;57:149-164.
33. Damopoulos D, Lerch TD, Schmaranzer F, et al. Segmentation of the proximal femur in radial MR scans using a random forest classifier and deformable model registration. International Journal of Computer Assisted Radiology and Surgery 2019;14(3):545-561.

34. Schauwecker N, Xi Y, Slepicka C, et al. Quantifying differences in femoral head and neck asphericity in CAM type femoroacetabular impingement and hip dysplasia versus controls using radial 3DCT imaging and volumetric segmentation. The British Journal of Radiology 2020;93(1110):20190039.
35. Frank JM, Lee S, McCormick FM, et al. Quantification and correlation of hip capsular volume to demographic and radiographic predictors. Knee Surgery, Sports Traumatology, Arthroscopy 2016;24(6):2009-2015.
36. Weinberg DS, Williamson DFK, Millis MB, Liu RW. Decreased and increased relative acetabular volume predict the development of osteoarthritis of the hip AN OSTEOLOGICAL REVIEW OF 1090 HIPS. Bone Joint J 2017;99B(4):432-439.
37. Audenaert EA, Mahieu P, Pattyn C. Three-Dimensional Assessment of Cam Engagement in Femoroacetabular Impingement. Arthroscopy: The Journal of Arthroscopic and Related Surgery 2011;27(2):167-171.
38. Kang RW, Yanke AB, Orias AE, Inoue N, Nho SJ. Emerging Ideas: Novel 3-D Quantification and Classification of Cam Lesions in Patients With Femoroacetabular Impingement. Clinical Orthopaedics and Related Research® 2013;471(2):358-362.
39. Guidetti M, Malloy P, Alter TD, et al. MRI-and CT-based Metrics for the Quantification of Arthroscopic Bone Resections in Femoroacetabular Impingement Syndrome. Journal of Orthopaedic Research® 2021.
40. Treece GM, Poole KES, Gee AH. Imaging the femoral cortex: thickness, density and mass from clinical CT. Medical image analysis 2012;16(5):952-965.
41. Mannava S, Geeslin AG, Frangiamore SJ, et al. Comprehensive Clinical Evaluation of Femoroacetabular Impingement: Part 2, Plain Radiography. Arthroscopy Techniques 2017;6(5):E2003-E2009.
42. Neubert A, Wilson KJ, Engstrom C, et al. Comparison of 3D bone models of the knee joint derived from CT and 3T MR imaging. European Journal of Radiology 2017;93:178-184.
43. Xia Y, Fripp J, Chandra SS, Walker D, Crozier S, Engstrom C. Automated 3d quantitative assessment and measurement of alpha angles from the femoral head-neck junction using mr imaging. Physics in Medicine and Biology 2015;60(19):7601-7616.
44. Arezoomand S, Lee WS, Rakhra KS, Beaule PE. A 3D active model framework for segmentation of proximal femur in MR images. International Journal of Computer Assisted Radiology and Surgery 2015;10(1):55-66.
45. Zeng G, Zheng G. Deep Volumetric Shape Learning for Semantic Segmentation of the Hip Joint from 3D MR Images. Computational Methods and Clinical Applications in Musculoskeletal Imaging. Cham: Springer International Publishing; 2019. p. 35-48.
46. Pham DD, Dovletov G, Warwas S, Landgraeber S, Jäger M, Pauli J. Deep segmentation refinement with result-dependent learning. Bildverarbeitung für die Medizin 2019: Springer; 2019. p. 49-54.

Automated cam FAI Morphology Analysis# Appendix A: Comparison of the current study with related works

**A.1.** Comparison of the present study with related works on the proximal femur, femoral head, and hip joint segmentation in MRI data.

| Study | Segmentation Description | Subjects | MRI strength (T) | Average DSI |
|---|---|---|---|---|
| 15 | Proximal femur: Multi-atlas method in 3D | 30 | 3.0 | 0.95 |
| 15 | Proximal femur: Active shape models in 3D | 30 | 3.0 | 0.946 |
| 12 | Femoral head bone: Focused Shape Models in 3D | 25 | 3.0 | 0.98 |
| 44 | Proximal femur: Parametric deformable model in 3D (Dataset 1) | 1 | 3.0 | 0.883 |
| 44 | Proximal femur: Parametric deformable model in 3D (Dataset 2) | 1 | 3.0 | 0.883 |
| 30 | Proximal femur: 3D U-net | 20 | - | 0.987 |
| 29 | Proximal femur: 2D and 3D U-net | 86 | 3.0 | 0.95 |
| 25 | Proximal femur: Multi-level latent shape space constrained 3D U-net | 25 | - | 0.954 |
| 26 | Proximal femur: 3D U-net | 20 (images) | | 0.987 |
| 45 | Proximal femur: Deep volumetric shape learning in 3D | 24 | - | 0.933 |
| 31 | Proximal femur: 3D tiled convolution in 3D | 25 (images) | - | 0.9814 |
| 32 | Proximal femur: Holistic decomposition convolution in 3D | 25 (images) | - | 0.9814 |
| 33 | Proximal femur: Random forest classifier with deformable model registration | 25 | 1.5 | 0.9637 |
| 46 | Proximal femur: Deep segmentation in 2D U-net | 6 | 1.5 | 0.8694 |
| 46 | Proximal femur: Deep segmentation in 2D Ref-Net | 6 | 1.5 | 0.8617 |
| 28 | Proximal femur: 2D CNN | 13 | 1.5 | 0.8973 |
| 27 | Femur bone: Resnet50-segnet | 38 | 1.5 | 0.907 |
| **Our Network: U-net results** | Proximal femur: 3D U-net | 97 | 3.0 | 0.958 |
| **FSM results** | | | | 0.964 |

**A.2.** Comparison of the femoral head volume results with related works. Where MRA = Magnetic Resonance Arthrography, CT = computed tomography, M = Male, F = Female, $\mu$ = mean, $\sigma$ = standard deviation

| Study | Method Description | Subjects | Image Modality | $\mu$ volume ($\sigma$) (cm$^3$) | | | Volume range (cm$^3$) | $\mu$ radii ($\sigma$) (mm) |
| --- | --- | --- | --- | --- | --- | --- | --- | --- |
| | | | | Male | Female | Sex Not Specified | | |
| 23 | Mimics segmentation; manual femoral head centre initialization; iterative search: $\sigma$ to point cloud minimization | M: F 69: 69 | CT | - | - | - | - | M: 25.4 (1.3) F: 22 (1.3) |
| 22 | Region growing segmentation | M: F 20: 23 | CT | 62.9 (10.8) | 41.8 (8.6) | - | 24.4 - 85.2 | - |
| 9 | Region growing segmentation | M: F 13: 14 | CT | - | - | 49.7 (11.5) | - | - |
| 34 **Abnormal FAI results** | Region growing segmentation | 79 | CT | - | - | 47.84 (9.65) | - | - |
| 35 | Manual Analysis | M: F 44: 53 | 1.5T MRA | 57.16 (9.71) | 37.98 (5.71) | - | - | - |
| 36 | Cadaveric Measurements | 1090 hips | - | - | - | 58.194 (11.998) | - | - |
| **Ours** | 3D U-net, Hough transform, spherical fitting | 97 | 3T MR | 66.12 (7.67) | 46.02 (6.83) | | M: 52.39 - 81.03 F: 34.31 - 59.49 | - |

**A.3.** Comparison of the cam morphology quantification results with related works. Where N = number of subjects, CT = computed tomography, SA = surface area, $\mu$ = mean, $\sigma$ = standard deviation, M = Male, F = Female, SNS = Sex Not Specified

| Study | Method Description | N | Image Modality | $\mu$ Volume ($\sigma$) (cm$^3$) | Volume range (cm$^3$) | $\mu$ SA ($\sigma$) (cm$^2$) | $\mu$ height ($\sigma$) (mm) | Max height ($\sigma$) (mm) |
|---|---|---|---|---|---|---|---|---|
| [37] | Mimics segmentation; femoral head and neck radii constraints | 7 (M) | CT | - | - | 3.735 (1.547) | | - |
| [38] | Mimics segmentation; femoral head radii constraints | 5 | CT | - | - | - | SNS: 8.26 | |
| [10] FAI Results | Manual segmentation; estimation of normal surface | 5 | 3T MRI | SNS: 0.188 (0.247) | | SNS: 0.52 (0.60) | SNS: 3.9 (2.1) | |
| [23] | Mimics segmentation; femoral head radii constraints | M: F 69: 69 | CT | M: 0.433 (0.471) F: 0.089 (0.124) | - | - | M: 1.51(0.75) F: 0.66(0.61) | |
| [22] | Region growing segmentation | M: F 20: 23 | CT | M: 6.7 (2.5) F: 4.3 (3.4) | 1.2 - 12.5 | - | | - |
| [9] | Region growing segmentation | M: F 13: 14 | CT | SNS: 4.6 (2.6) | | - | | - |
| [34] Abnormal FAI Results | Region growing segmentation | 79 | CT | SNS: 7.96 (2.78) | - | - | | - |
| [39] MR results | Mimics segmentation | 7 | CT and MR | SNS: 0.940 (0.537) | - | SNS: 6.158 (2.324) | SNS: 1.6 (0.4) | SNS: 3.7 (0.9) |
| Ours | 3D U-net and FSM | M: F 56: 41 | 3T MR | M: 1.136 (0.659) F: 0.338(0.280) | M: 0.022 - 3.002 F: 0 - 1.164 | M: 6.574 (2.030) F: 3.069 (1.756) | M: 1.94 (0.86) F: 1.00 (0.57) | M: 3.89 (1.51) F: 2.23 (1.09) |

**A.4.** Correlation between femoral head volume and cam morphology volume in male (panel A) and female (panel B) participants.

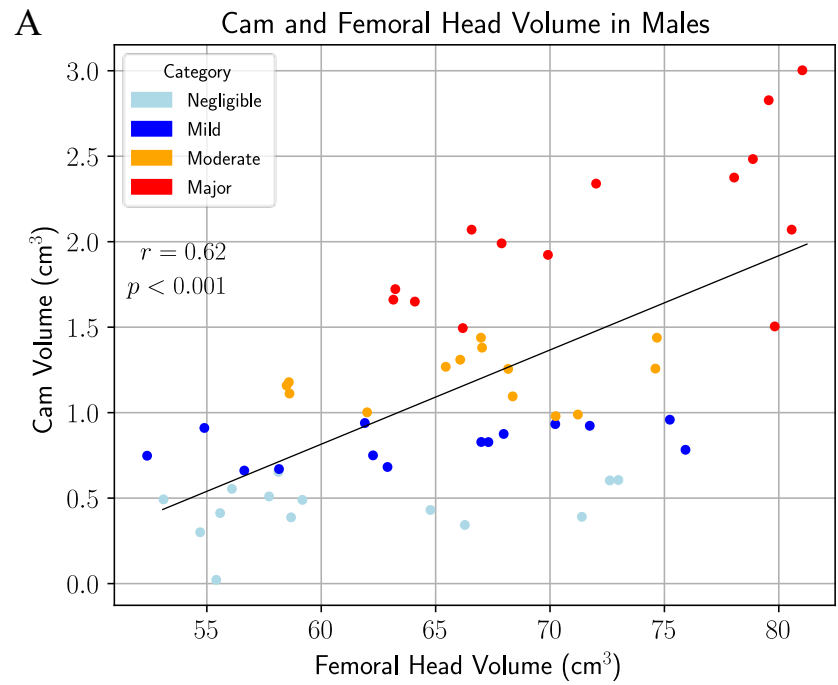
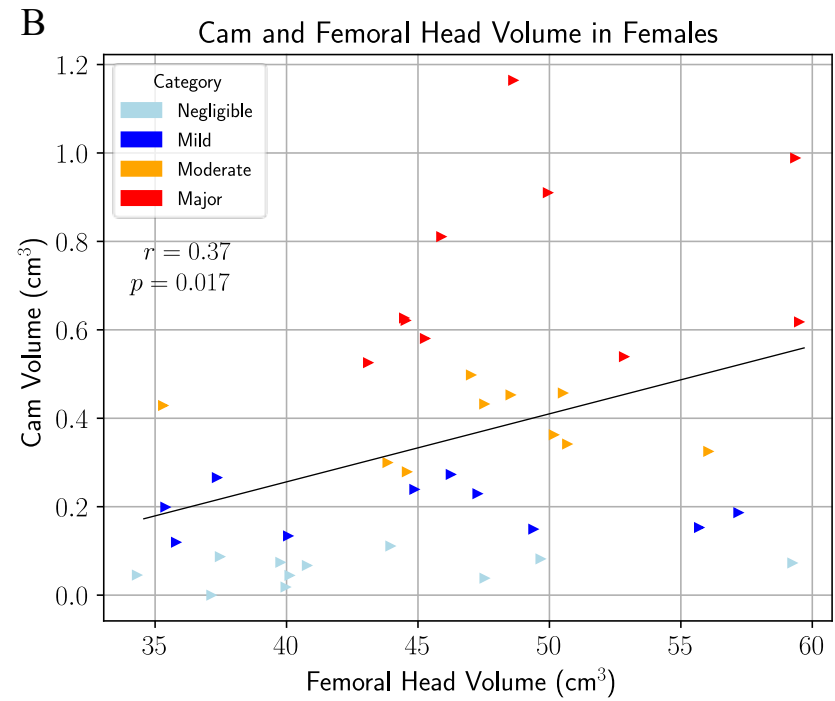